\documentclass[twocolumn,amsmath,amssymb,prd]{revtex4}

\usepackage{graphicx}
\usepackage{rotating}
\def\arcdeg{$^{\circ}$}

\def\sun{\odot}
\def\aaps{A\&AS}
\def\aap{A\&A}
\def\mnras{MNRAS}
\def\aj{AJ}
\def\araa{ARA\&A}
\def\pasp{PASP}
\def\apjl{ApJL}

\begin{document}

\title{Light-Element Abundances of Giant Stars in the Globular Cluster M71 (NGC 6838)}

\author{M.~J. Cordero\footnote{Department of Astronomy, Indiana University,
Swain West 319, 727 East Third Street, Bloomington, IN 47405--7105, USA}
\footnote{Visiting Astronomer, Kitt Peak National
Observatory, National Optical Astronomy Observatories, which is
operated by the Association of Universities for Research in
Astronomy, Inc. (AURA) under cooperative agreement with the
National Science Foundation.}}
\email{mjcorde@lsw.uni-heidelberg.de}
\affiliation{Zentrum f\"ur Astronomie der Universit\"at Heidelberg, Landessternwarte, K\"onigstuhl 12, Heidelberg, Germany}

\author{C. A. Pilachowski}
\email{catyp@astro.indiana.edu}
\affiliation{Astronomy Department, Indiana University Bloomington, Swain West 319, 727 E. 3rd Street, Bloomington, IN 47405-7105, USA}

\author{C.~I. Johnson\footnote{Clay Fellow}}
\email{cjohnson@cfa.harvard.edu}
\affiliation{Harvard-Smithsonian Center for Astrophysics, 60 Garden Street, MS-15, Cambridge, MA 02138, USA}


\author{E. Vesperini}
\email{evesperi@indiana.edu}
\affiliation{Astronomy Department, Indiana University Bloomington, Swain West 319, 727 E. 3rd Street, Bloomington, IN 47405-7105, USA}

\begin{abstract}
Aluminum is the heaviest light element displaying large
star--to--star variations in Galactic globular clusters (GCs).
This element may provide additional insight into the origin of the
multiple populations now known to be a common place in GCs, and
also the nature of the first-generation stars responsible for a
cluster's chemical inhomogeneities. In a previous analysis we
found that, unlike more metal-poor GCs, 47 Tuc did not exhibit a
strong Na-Al correlation, which motivates a careful study of the
similar metallicity but less massive GC M71.  We present chemical
abundances of O, Na, Al, and Fe for 33 giants in M71 using spectra
obtained with the WIYN-Hydra spectrograph. Our spectroscopic
analysis finds that, similar to 47 Tuc and in contrast with more
metal-poor GCs, M71 stars do not exhibit a strong Na-Al
correlation and span a relatively narrow range in [Al/Fe], which are characteristics that GC formation models must reproduce.
\end{abstract}

\maketitle

\section{Introduction}

Multiple and discrete photometric sequences found in most globular
clusters (GCs), e.g. Omega Cen (e.g. Bedin et al. 2004), M54 (e.g.
Bellazzini et al. 2008), NGC 2808 (e.g. Bedin et al. 2000, Piotto
et al. 2007), NGC 1851 (e.g. Milone et al. 2008, Han et al. 2009),
and NGC 6752 (e.g. Milone et al. 2009), demonstrate that GCs are
not simple stellar populations (see also Piotto et al. 2014 for the first results of a large HST UV survey of Galactic GCs). 
In addition to photometric studies, the large star-to-star light-element abundance variations
of C, N, O, Na, Mg, and Al, coupled with characteristic
(anti)correlations (e.g., C-N, Na-O, Mg-Al) extending down to the
main-sequence (MS), are considered compelling evidence that GCs
host multiple stellar populations with unique chemistry (e.g., see
reviews by Gratton et al. 2004; 2012). In this framework a first
generation of stars forms with a composition matching the
cluster's primordial gas, and later the more massive stars inject
processed material into the cluster interstellar medium.
Subsequent generations that form then reflect a composition that
is a mixture of the primordial gas cloud and any retained
processed material. Rapidly rotating massive stars (e.g., Decressin et al. 2007, Decressin et al. 2010; Charbonnel et al. 2013), asymptotic giant
branch (AGB) stars (e.g., Ventura et al. 2001, D'Ercole et al. 2008; D'Antona et al.
2012), massive binary stars (de Mink et al. 2009), and
supermassive stars (Denissenkov \& Hartwick 2014) have all been
considered to explain the abundances patterns of second generation
stars in GCs. Improving our understanding of the chemical
enrichment history of GCs and the nature of the progenitors
responsible for the gas pollution requires increasing both the
number of stars with known chemical abundances and the number of
clusters surveyed sampling a broad range in both GC mass and
metallicity. For instance, Carretta et al. (2010) found that some
of the more luminous or massive GCs display a more extended Na-O
anticorrelation. Furthermore, larger spreads in Al content are
found in more metal-poor GCs, such as M13 (e.g., Shetrone 1996,
Kraft et al. 1997, Cavallo \& Nagar 2000, Sneden et al. 2004;
Cohen \& Mel\'endez 2005; Johnson et al. 2005), M5 (Ivans et al.
2001), M80 (Cavallo et al. 2004), NGC 6752 (Carretta et al. 2009b;
hereafter CBG09b), suggesting that the Al-production is regulated
by the cluster's metallicity. Moreover, an Al-Na correlation,
which is typically found in more metal-poor GCs such as M13 and M3
(e.g. Cavallo and Nagar 2000), is absent in the more metal-rich
populations of Omega Cen (Johnson \& Pilachowski 2010) and 47 Tuc
(Cordero et al. 2014).

Multiple lines of spectroscopic evidence support the multiple
populations scenario in M71. Cohen (1999) reported an
anticorrelation between CN and CH, which trace N and C,
respectively, and a bimodal CN distribution for a sample of MS
stars. Moreover, Briley et al. (2001) found that red giant
branch (RGB) stars exhibit the same scatter in C and N and CN
bimodality as MS stars, which lead the authors conclude that these
chemical inhomogeneities detected in M71 were imprinted at birth
and that variations due to deep mixing during the first dredge-up
are less important at this metallicity. Another piece of
evidence revealing that deep internal mixing in more metal-rich
GCs plays a minor role, if any, in shaping the light-element
abundances pattern described by red giants comes from the study of
Briley et al. (2001), which used DDO photometry of M71 giants to
trace CN and CH band strengths. Their study demonstrated that the
scatter in C and N exhibited by less evolved giants is comparable
to the one found in the more luminous giants. The missing
dependence of this chemical feature with luminosity is also found
in 47 Tuc giants, a GC with similar metallicity, as shown in Fig.
3 from Briley et al. (2001) using results from Norris \& Freeman
(1979) and Briley (1997). 

Regarding the characterization of other light elements involved in proton-capture
reactions, Ram\'irez \& Cohen (2002) reported Na and O abundances
for a sample of 25 stars in M71 covering a large luminosity range,
finding that the Na-O anticorrelation extends down to the
main-sequence turn-off. This result demonstrated that the origin
of the Na-O anticorrelation in M71 is linked to a self-pollution
scenario, since less evolved stars do not have temperatures high
enough to activate the ON and NeNa cycles. Moreover, from a sample
of nine stars, Mel\'endez \& Cohen (2009; hereafter MC09) found
two groups in M71, one resembling the light element abundances
found in halo field stars at that metallicity and a second
population of O-depleted, Na/Al enhanced stars with higher
$^{26}$Mg/Mg ratios consistent with AGB pollution. Furthermore,
previous chemical studies of M71 indicate that the cluster
exhibits a more narrow range in [O/Fe] than do more metal-poor
clusters and to date extreme O-depleted stars ([O/Fe]$<-0.4$) have
not been found in this cluster (Sneden et al. 1994; Ram\'irez \&
Cohen 2002; MC09; and Carretta et al. 2009a, hereafter CBG09a).

In order to gather more information in the more metal-rich regime
of Galactic GCs we determined O, Na, and Al abundances in 47 Tuc
(Cordero et al. 2014). In the present work, we extend the study of
light-element abundances to another old and metal-rich GC, M71,
with the goals of: i)  increasing the small dataset of abundances
available for this cluster; ii) assessing the degree of Al
enhancement and its dispersion within the cluster; and iii)
exploring whether an Na-Al correlation is present in M71.
Furthermore, the total mass and metallicity of GCs are often
correlated with their horizonal branch morphology as well as light
element chemistry (Carretta et al. 2010), and since M71 and 47 Tuc
have similar [Fe/H] but differ significantly in their present-day
mass, a comparison between the two clusters may provide insight
into possible differences in their formation mechanisms. The paper
is organized as follows. A description of the observations, the
analysis, and a comparison to the literature are presented in
sections 2 through 4. A discussion of our results and our
conclusions are given in sections 5 and 6, respectively.

\section{Observations and Data Reduction}

\subsection{Observations}

We observed red giant branch stars in the GC M71 using the Hydra
multi-fiber positioner and bench spectrograph on the WIYN
telescope located at Kitt Peak National Observatory in June 2000,
June 2013, and July 2013. Different fiber configurations were used
during each run because only 13-17 stars can be observed on a
given setup due to the small angular diameter of the cluster and
the fiber size/placement tolerances. The remaining fibers in each
configuration were used to measure the background sky spectra.
Table \ref{table_log} shows the dates of observations and
wavelength coverage of each Hydra configuration. For all Hydra
runs the spectrograph was fed with the 200 $\mu$m ``red'' fiber
cable and the Bench Spectrograph Red Camera was used. The 2001
observations employed the 316 l/mm (63.4\arcdeg) echelle grating,
the former SITE 2048 $\times$ 2048 pixel CCD (t2kc), and the
filter X19 which yielded spectra ranging from
$\lambda\lambda$6470-6860 \AA, at a dispersion of 0.20 \AA\
pixel$^{-1}$. The 2013 observations used the 316 l/mm
(63.4\arcdeg) echelle grating, the SITE 2600 $\times$ 4000 pixel
CCD (STA1), and the filter X18 which covered the spectral range
from $\lambda\lambda$6100-6350 \AA, at a dispersion of 0.16 \AA\
pixel$^{-1}$.

\subsection{Selection of Stars}
\begin{figure}
\includegraphics[width=0.5\textwidth]{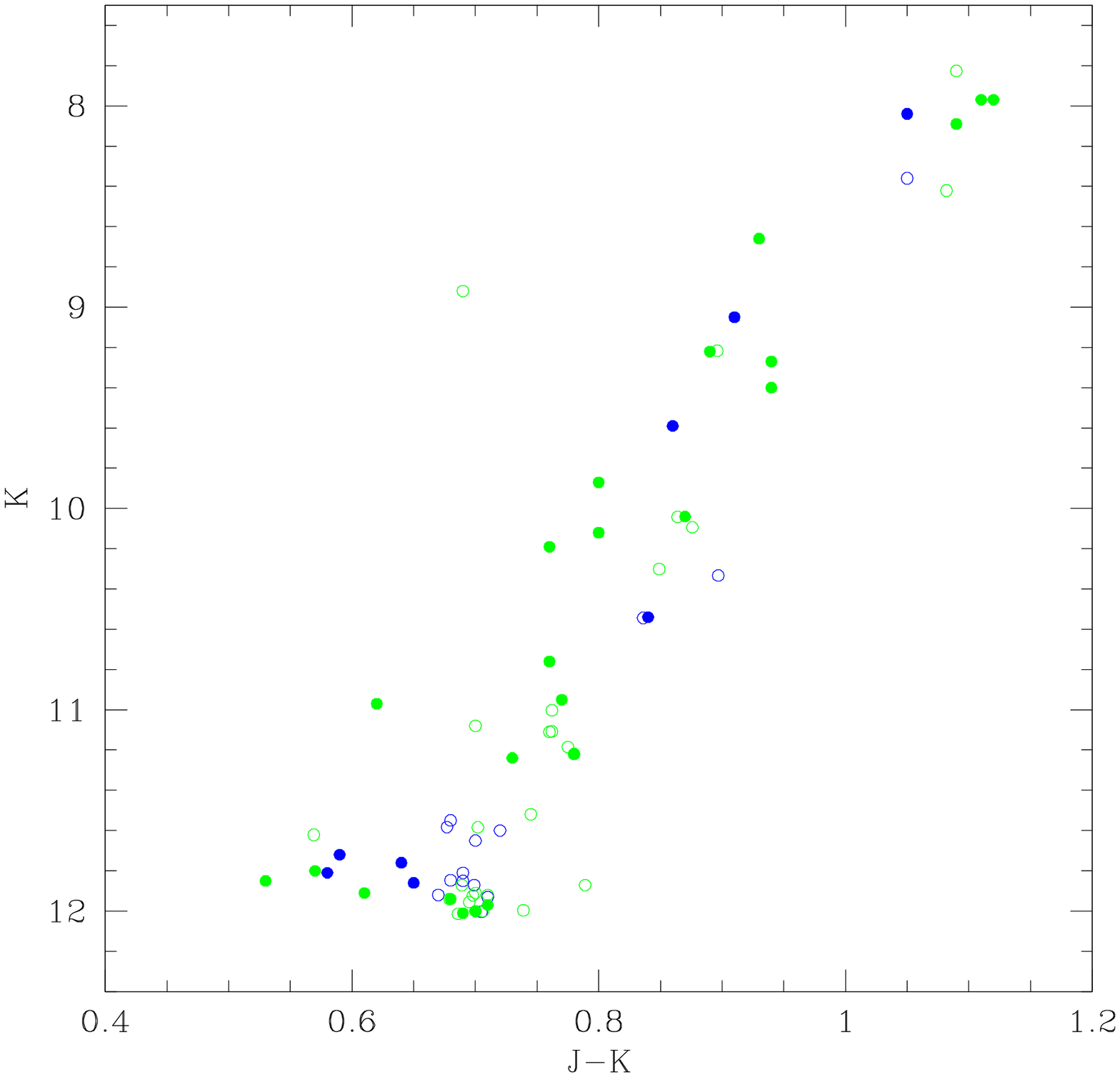}
\caption{K$_{\rm S}$ vs.
J-K$_{\rm S}$ color magnitude diagram of stars observed in 47 Tuc.
The stars appear color-coded by population assignment using the
criteria defined by Carretta et al. (2009b). Stars assigned to the
primordial population are shown in blue and stars assigned to the
intermediate population are shown in green (see section 5.1 for
the definitions of population groups).  Our Hydra sample is
represented by filled circles while stars from Carretta et al.
(2009b) are shown with open symbols.  J and K$_{\rm S}$ magnitudes
were obtained from the 2MASS Point Source Catalog.}
\label{fig_CMD}
\end{figure}

Coordinates and membership probabilities were obtained from the
proper motion survey of Cudworth (1985). We targeted only stars
with membership probabilities greater than 80\%, and note that
only seven stars in our sample have membership probability below
90\%. Furthermore, to confirm membership radial velocities were
measured using the IRAF task {\it fxcor} which cross-correlates
the observed spectra with synthetic spectra of similar atmospheric
parameters and resolution. This procedure resulted in an average
heliocentric velocity of -22.5 km s$^{-1}$ with $\sigma$=2.6 km
s$^{-1}$ and typical uncertainties of $\sim$ 0.4 km s$^{-1}$, in
agreement with previous studies (i.e., Peterson \& Latham 1986,
Cohen et al. 2001). The stars included in this study range in
magnitude from 12.1 $\leq$ V $\leq$ 14.9, from near the RGB-tip to
slightly below the HB. The final sample includes spectra for 25
stars from the 2001 observations, and 33 stars from the 2013
observations, with 25 stars observed in both wavelength regions.
Stellar identifications and V magnitudes were taken from Cudworth
(1985) and near-infrared photometry (J and K$_{\rm S}$) was taken
from the 2MASS Point Source Catalog (Skrutskie et al. 2006). A K$_{\rm S}$
vs J-K$_{\rm S}$ color-magnitude diagram of stars observed in M71
is shown in Fig. \ref{fig_CMD}, including all our Hydra sample and
the sample from CBG09b.

\subsection{Data Reduction}

Calibration data for both Hydra runs included ThAr comparison
lamp, bias, and flat field frames on all nights. Furthermore, a
set of dark frames was taken on the 2013 observing run. Basic data
reductions were carried out using IRAF\footnote{IRAF is
distributed by the National Optical Astronomy Observatories, which
are operated by the Association of Universities for Research in
Astronomy, Inc., under cooperative agreement with the National
Science Foundation."} routines. The {\it ccdproc} task was used to
trim the overscan region and to subtract the bias and dark levels.
The {\it dohydra} script was used to trace the spectra, to apply a
flat-field correction, to correct for scattered light, to extract
the one-dimensional spectra, to apply the wavelength calibration,
and to subtract the background sky spectra. The continuum for each
spectrum was normalized using a low order cubic spline. Standard stars for removing telluric absorption features near the [O I] line at 6300 were not obtained as part of the original data set, limiting the determination of an oxygen abundance to just a subset of our stars. Weak telluric features are present in our spectra and are blended with the oxygen line. However, we were able to convolve the high resolution telluric spectrum provided with the Arcturus Atlas (ftp://ftp.noao.edu/catalogs/arcturusatla/visual/) to match the resolution of our data and successfully divide out the weak telluric absorption lines in our spectra. With the telluric lines removed, we were able to determine oxygen abundances for 22 giants in our sample. Finally, the spectra were co-added using the {\it scombine} task after
applying the appropriate wavelength shift. The final spectra have
signal-to-noise ratios of typically 70-200 per pixel.

\section{Abundance Analysis and Uncertainties}

\subsection{Derivation of Abundances}
Chemical abundances of M71 giants were determined using the 2010
version of the code MOOG (Sneden 1973). LTE model stellar
atmospheres were interpolated, with an adopted metallicity of
[M/H]=--0.80\footnote{We use the standard spectroscopic notation
where [A/B]$\equiv$log(N$_{\rm A}$/N$_{\rm B}$)$_{\rm
star}$--log(N$_{\rm A}$/N$_{\rm B}$)$_{\sun}$ and log
$\epsilon$(A)$\equiv$log(N$_{\rm A}$/N$_{\rm H}$)+12.0 for
elements A and B.}, from the $\alpha$-enhanced ATLAS9 grid of
non-overshoot models (Castelli \& Kurucz 2003).

Initial effective temperatures ($T_{\mbox{eff}}$) were obtained
from the infrared flux method using the analytic functions
presented in Alonso et al. (1999) and the color transformations
summarized in Johnson et al. (2005). We used the reddening value
E(B-V)=0.25 recommended by Harris (1996; 2010 edition). We derived surface gravities from the
absolute bolometric magnitudes and effective temperatures. We used
the bolometric corrections described in Alonso et al. (1999), a
distance modulus of (m-M)$_{V}$ = 13.80 (Harris 1996, 2010
edition), a bolometric magnitude of 4.74 for the Sun , and a mass of 0.8 $M_{\odot}$ was assumed for all the
stars. The limited spectral coverage prevents derivation of the
surface gravities from the ionization equilibrium of Fe I and Fe
II. An initial microturbulence value was obtained using the
v$_{t}$ relation given by Johnson et al. (2008), and then adjusted
to minimize the dependence of derived [Fe/H] on reduced equivalent
width ($\log W/\lambda$). The final adopted atmospheric parameters
are given in Table \ref{tabshort}, which also includes,
abundances, reference population assignment, distance from the
cluster center in arcmin, and heliocentric radial velocities
obtained in the present work, in addition to stellar
identifications and membership probabilities from Cudworth (1985).

Iron abundances were obtained from equivalent width measurements
and the line list from Johnson \& Pilachowski (2010). The IRAF
{\it splot}  package was used to measure equivalent widths
interactively and only lines with log W/$\lambda$ $<$--4.5 were
used for the abundance analysis. We used typically about 27 Fe I
lines. Solar reference abundances used in this study can be found
in Johnson \& Pilachowski (2010).

\begin{figure}
\includegraphics[width=0.5\textwidth]{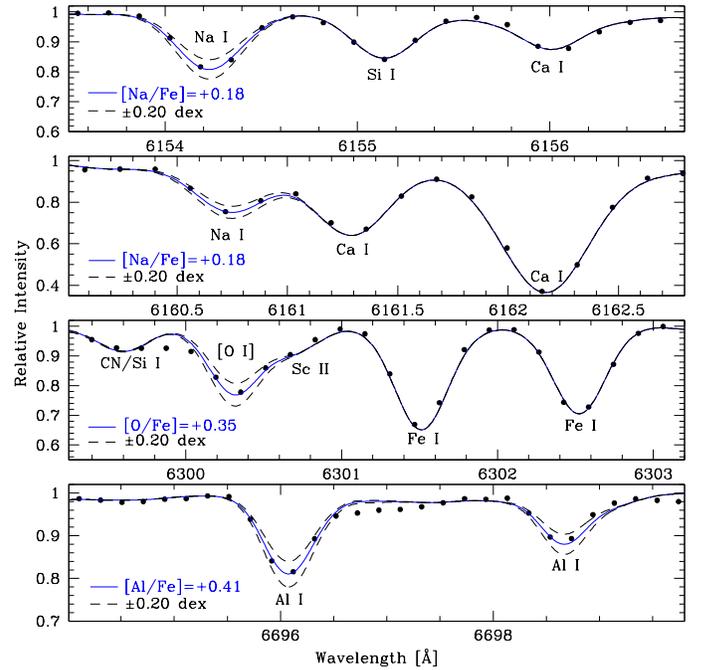}
\caption{Sample spectra of star
A4 showing the determination of the sodium, oxygen, and aluminum
abundances using spectrum synthesis.  The observed spectrum is
shown as dots, while the synthetic spectra are shown as dashed
lines for three different abundances.  The adopted abundance is
shown as a solid blue line.  Differences of 0.2 dex can be easily
discerned.} \label{fig_synth}
\end{figure}

\begin{figure}
\includegraphics[width=0.5\textwidth]{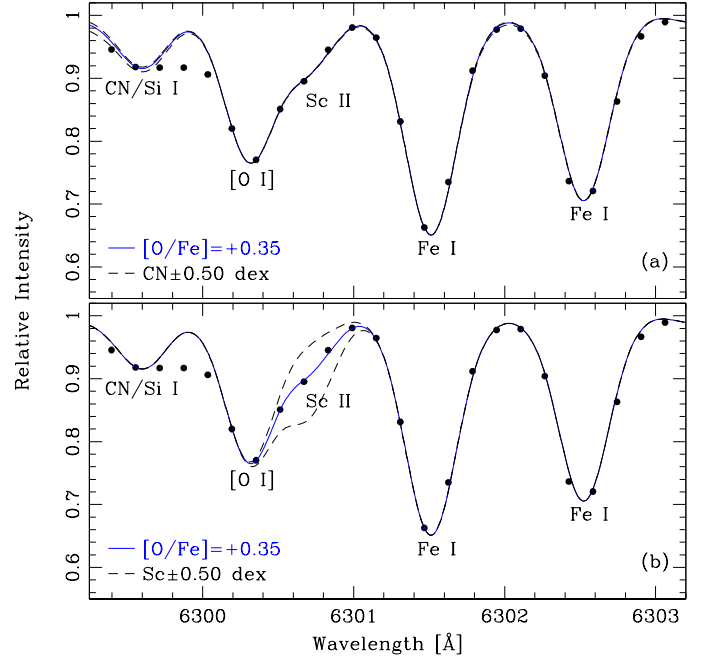}
\caption{Spectrum synthesis
around the oxygen line showing the effect of altering the
abundances of blended features in star A4. The upper panel shows
that the oxygen abundance is unaffected by a change in CN
abundance of $\pm$ 0.50 dex (dashed lines).  The lower panel shows
that the Sc feature must be included in the synthesis to derive a
reliable oxygen abundance.} \label{fig_CNSc}
\end{figure}

Abundances of the light-elements oxygen ([O I] $\lambda$6300 \AA),
sodium (Na I 6154, 6160 \AA\ doublet), and aluminum
($\lambda\lambda$ 6696, 6698 \AA\ Al doublet) were determined by
spectrum synthesis, which provides a more reliable abundance at
Hydra's resolution due to blending with other atomic and molecular
features. For instance, to obtain a good fit to the [O I]
$\lambda$6300 \AA\ line profile the Sc abundance must be adjusted
as shown in Fig. \ref{fig_CNSc}. Typically, a [Sc/Fe] ranging from 0.05 to 0.15 resulted in a well fit line profile, which is consistent with published Sc values published by Ramirex \& Cohen (2002) (-0.39 to 0.29). Furthermore, the [O I]
$\lambda$6300 \AA\ line is blended with Ni. The Na $\lambda$6160
\AA\ feature is blended with Ca and the Al $\lambda\lambda$ 6696,
6698 \AA\ doublet is blended with Fe and CN. The synthesis line
lists used in this study were the same as those used in Johnson \&
Pilachowski (2012). The synthetic spectra were convolved with a
Gaussian smoothing kernel to match the resolution of unblended
lines. An example of synthesis around each spectral feature is
presented in Fig. \ref{fig_synth}.

Given the uncertain nature of Na NLTE corrections, we do not
include such corrections here. However, departures from LTE
decrease with increasing metallicity and are expected to be
$\lesssim$ 0.1 dex for M71 red giants (Lind et al. 2011).

\subsection {Uncertainties}

Photometric temperatures based on the empirical relations for
V-K$_{\rm S}$ and B-V colors extracted from Alonso et al. (1999)
have a 1$\sigma$ value ranging from 20-96K. Additionally, we find
in a comparison of our derived $T_{\mbox{eff}}$ values with those
of CBG09b and MC09 average differences of 80K and 37 K,
respectively. Therefore, we adopt 100 K as a conservative level of
uncertainty in the derived $T_{\mbox{eff}}$ for each star.
Similarly, comparing our surface gravities we found a
1$\sigma$-scatter of 0.06 and 0.20 dex with these previous
studies, and adopted the highest value as a reasonable uncertainty
in our log g values. Changes in microturbulence velocities of up
to 0.10 km s$^{-1}$ resulted in a lack of trend of the Fe
abundance with respect to the line strength or reduced equivalent
width. However, we find a 1$\sigma$-scatter of 0.13 and 0.17 km
s$^{-1}$ with the studies mentioned above; to be on the
conservative side we adopted an uncertainty of 0.2 km s$^{-1}$ in
our microturbulence values. An assessment of the abundance ratios
sensitivity to the uncertainties in the atmospheric parameters is
performed by generating a grid of atmospheres while changing one
parameter a time by its uncertainty. Then each abundance change
introduced by varying the atmospheric parameters is added in
quadrature in addition to the line-to-line dispersion to estimate
the total uncertainty in the abundance. We found that the
abundances of each pair of features for Na and Al agree within
0.10 dex or less, which translates into a line-to-line dispersion
$\leq$ 0.07.  In the case of oxygen where we only have a single
line, we adopted a measurement uncertainty of 0.07 dex. To
estimate the uncertainties introduced by spectrum synthesis, we
applied small changes in the continuum placement, smoothing
factor, and the abundance values while still obtaining a
reasonable fit to the observed spectra. These procedure yielded an
uncertainty $\leq$ 0.05 dex depending on the S/N ratio of the
observed spectrum. We used this value to estimate the total error
for Na and Al when the line-to-line dispersion was smaller. Table
\ref{table_sensitivities} shows derived uncertainties for two
stars with different atmospheric parameters. We find that each
element has typical uncertainties ranging from 0.10-0.15 dex.

\section{Results}

\begin{figure}
\includegraphics[width=0.5\textwidth]{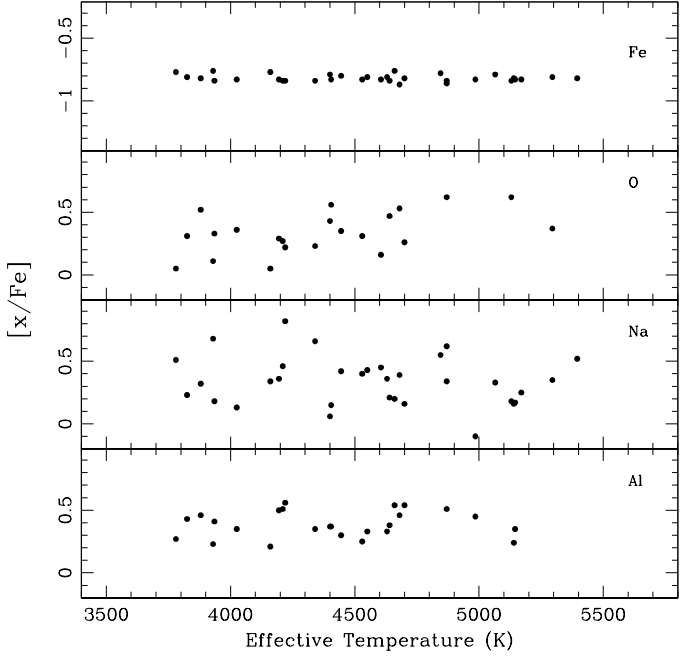}
\caption{Abundance ratios as a
function of the effective temperature.} \label{fig_teff}
\end{figure}

The Fe abundances in our sample of 33 stars cover a range of
$\sim$0.10 dex, consistent with typical systematic uncertainties
(0.12 dex). We find an average value of [Fe/H]=--0.82 and a
standard deviation of 0.03. Thus, the M71 metallicity is
remarkably similar to that of 47 Tuc (e.g. [Fe/H]=--0.79 from
Cordero et al. 2014). Moreover, our Fe abundances agree well with
previous studies: [Fe/H]=--0.79 (Sneden et al. 1994),
[Fe/H]=--0.80 (Boesgaard et al. 2005), [Fe/H]=--0.79 (MC09), and
[Fe/H]=--0.81 (CBG09a).  Our derived abundances as a function of
effective temperatures are presented in Fig. \ref{fig_teff}.

Oxygen abundances were measured for 22 stars after removing telluric contamination affecting the
profile of the [O I] $\lambda$ 6300 \AA\ line. We found an average
[O/Fe]=0.4 dex ($\sigma$=0.13), covering a short range ($\sim$0.4
dex). Sneden et al. (1994) analyzed the spectra of ten giants,
finding a very narrow range for O ($\sim$0.2 dex), with [O/Fe]
ratios above solar for all the stars, and an average
[O/Fe]$\sim$0.3 dex. MC09 and CBG09a also reported a small range
in their O abundances ($\sim$0.3-0.4 dex), supersolar [O/Fe]
ratios, and a slightly higher average [O/Fe]$\sim$0.4 dex compared
to Sneden, in a sample of 9 and 31 RGB stars, respectively. A lack
of extreme O-depleted stars in M71 is consistent with the
correlation between the total mass of GCs and the extention of
their Na-O (CBG09a).

\begin{figure}
\includegraphics[width=0.5\textwidth]{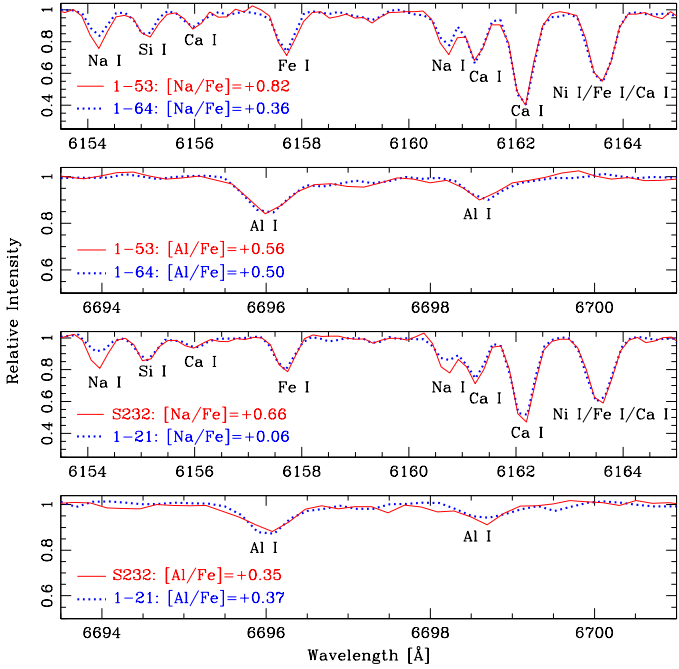}
\caption{Two pairs of stars with
identical atmospheric parameters exhibit differences in their Na
line strengths but have similar Al line strengths. The atmospheric
parameters for these stars are ($T_{\mbox{eff}}/\log(g)/v_t$):
4220/1.30/1.65 (1-53), 4195/1.34/1.60 (1-64), 4400/1.44/1.8
(1-21), 4340/1.62/1.55 (S232).} \label{fig_strengths}
\end{figure}

Na abundances exhibit the largest star-to-star variations in our
sample of 33 stars, exhibiting a broad range of values ($\sim$0.9
dex), and an average [Na/Fe]=0.34 dex ($\sigma$=0.20). We compared
our Na measurements with the results of CBG09a uncorrected for
NLTE effects and found no systematic offset using a common
subsample of seven stars. Large differences in Na content are also
found in pairs of stars that have similar atmospheric parameters,
as shown in Fig. \ref{fig_strengths}, which provides direct
evidence of star-to-star Na variations.

To separate our sample into primordial and intermediate
populations we adopted the criteria used by CBG09a. After removing
the NLTE correction used by CBG09a, the primordial population is
defined as those stars with [Na/Fe]$<$+0.20 dex, consistent with
the abundance ratios of field stars with similar metallicity. We
have seven stars in common with the sample presented by CBG09a,
which were used to compare the atmospheric parameters and
abundances. The average differences in the sense us minus CBG09a
are $\Delta(T_{\mbox{eff}})=-41$ K ($\sigma=80$ K), $\Delta(log
g)=-0.09$ cgs ($\sigma=0.06$ cgs), $\Delta(v_t)=0.00 $ km s$^{-1}$
($\sigma=0.17$ km s$^{-1}$), $\Delta([Fe/H])=0.05$ dex
($\sigma=0.06$ dex)), $\Delta([O/Fe])=-0.05$ dex ($\sigma=0.16$
dex, based on three stars), $\Delta$([Na/Fe])=0.00 dex
($\sigma=0.14$ dex, based on seven stars after removing their NLTE
correction). Our atmospheric parameters agree with CBG09a's and
the standard deviations are consistent with their typical
uncertainties. No systematic offset in [Na/Fe] is needed to place
CBG09a's sample onto our baseline and we decided not to shift the
oxygen abundances, since only three stars can be used as a
reference.

Al abundances were determined for 25 stars in our sample and cover
a range of $\sim$0.35 dex. We found an average [Al/Fe]=0.39
($\sigma$=0.11) which is remarkably similar to 47 Tuc (Cordero et
al. 2014). Published Al abundances for M71 are scarce. We compare
our results with MC09 and CBG09b which used samples of 9 and 11
stars, respectively. From a common subsample of five stars with
CBG09b we found an offset of --0.15 dex in the [Al/Fe] ratio (ours
minus theirs), while an offset of +0.18 dex was found with respect
to the study of MC09 from a common subsample of seven stars. The
bulk part of CBG09b's higher [Al/Fe] ratios can be explained by
their different reference solar abundance, which increases the
Al-to-Fe ratios by 0.22 dex. Moreover, the average differences in
atmospheric parameters between us and CBG09b for the five stars in
common are $\Delta(T_{\mbox{eff}})=-79$ K ($\sigma=54$ K),
$\Delta(log g)=-0.11$ cgs ($\sigma=0.07$ cgs), $\Delta(v_t)=-0.05
$ km s$^{-1}$ ($\sigma=0.16$ km s$^{-1}$), $\Delta([Fe/H])=0.05$
dex ($\sigma=0.08$ dex). Thus, we find that differences in the
atmospheric parameters between CBG09b and our analysis can account
for a change in the [Al/Fe] ratio of no more than 0.1 dex. The
average difference with respect to MC09 for the seven stars in
common are $\Delta(T_{\mbox{eff}})=+79$ K ($\sigma=37$ K),
$\Delta(log g)=0.05$ cgs ($\sigma=0.20$ cgs), $\Delta(v_t)=0.04 $
km s$^{-1}$ ($\sigma=0.13$ km s$^{-1}$), $\Delta([Fe/H])=-0.02$
dex ($\sigma=0.06$ dex). Thus, their lower [Al/Fe] ratios can be
explained, at least to the $\leq$0.1 dex level, by their lower
temperatures. Unfortunately, the solar reference scale of MC09 is
not available to assess whether it creates an offset with our
results.

\section{Discussion}

\subsection{Na-O Anticorrelation and Radial Distributions}

\begin{figure}
\includegraphics[width=0.5\textwidth]{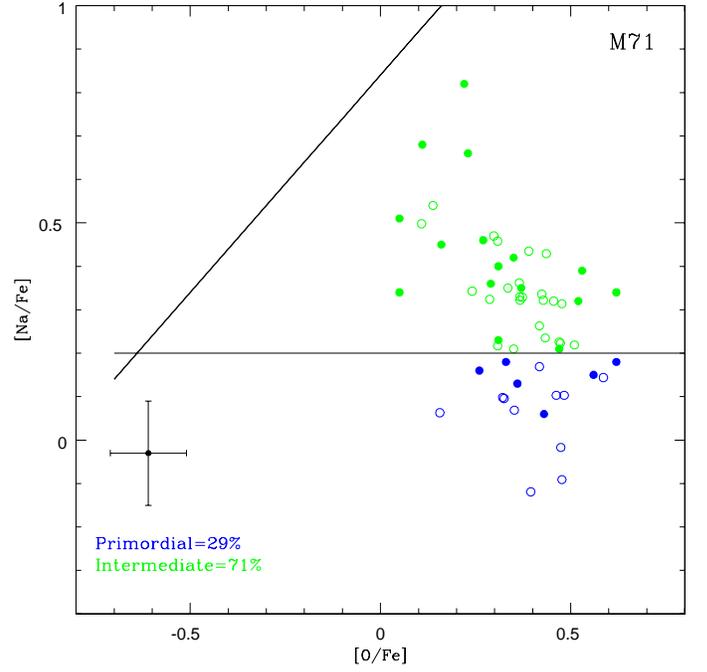}
\caption{The Na-O
anticorrelation.  Stars assigned to the primordial population are
shown in blue and stars assigned to the intermediate population
are shown in green. Our Hydra sample is represented by filled
circles, while those from Carretta et al. (2009b) are shown with
open symbols.  Stars are assigned to the primordial and
intermediate populations as defined in the text and indicated by
solid lines in the figure.} \label{fig_NaO}
\end{figure}

Determining the extent of the Na-O anticorrelation to low oxygen abundance is
of particular interest for understanding how GCs form. The
distribution of Na and O abundances for our sample follows the
same distribution on the Na-O plane as CBG09a's sample. For this
combined sample of 75 stars we find that the second generation
comprises 71$\%$ of the stars in M71, consistent with the range of
values obtained by CBG09a from a homogenous sample of 1500 stars
selected from 19 GCs ($\sim$ 50-70$\%$). Both samples lack extreme
oxygen-depleted stars, as shown in Fig. \ref{fig_NaO}. The absence
of stars with [O/Fe]$<$-0.5 dex in M71 may be explained by the
higher [O/Fe] yields of intermediate and massive stars at
[Fe/H]$\gtrsim$-1 (e.g., Ventura \& D'Antona 2009).Thus, the
less extreme O-depletion detected in M71 could be linked to a
lower efficiency of ON cycle reactions for polluters in this
metallicity regime. Moreover, the distribution of stars on the
Na-O plane is significantly less extended in oxygen for M71
compared to 47 Tuc (e.g. CBG09a, Cordero et al. 2014). Since M71
and 47 Tuc have similar metallicities, one might consider that the
same type of polluters are responsible for the chemical signatures
imprinted in the Na-rich/O-poor stars of these two clusters.
Carretta et al. (2010) noticed that less massive and/or more
metal-rich GCs span a shorter range in oxygen abundance and less
extended HB. Therefore, the lack of an extremely O-depleted
population can be a consequence of M71's lower total mass.
However, one of the lessons learned from the CBG09a survey is that extreme O-depleted stars are scarce.  Therefore, the samples considered so far could be too sparse to reveal the extention of the Na-O anticorrelation in M71. In any case, larger samples of oxygen
abundances are needed to effectively probe the tails of the O
distribution in M71.

Both spectroscopic (Cordero et al. 2014) and photometric (Milone
et al. 2012, Richer et al. 2013) studies have shown that 47 Tuc's
intermediate population is more centrally concentrated than its
primordial population. According to the Kolmogorov-Smirnov
statistical test (p-value=0.23) (\ref{fig_cum}), the spatial
distributions of the two populations found in M71 in this study
are not statistically different. Future photometric studies of
larger samples using UV filters sensitive to CN variations might
clarify whether there is indeed no radial variation in the
fraction of M71 intermediate population stars or the two
populations are characterized by a weak radial gradient that is
not detectable with the number of stars in our sample. In any
case, our study (Fig. \ref{fig_cum}) suggests that the two
populations in M71 are more spatially mixed than those of 47 Tuc.
A possible clue as to why 47 Tuc and M71 exhibit different degrees
of radial population mixing may be found in a comparison of the
clusters' dynamical ages, defined as the ratio of the cluster age
to its half-mass relaxation time. Assuming the clusters have
similar ages (e.g. 12 Gyr), the dynamical age of 47 Tuc is about
$t/t_{rh} \sim 3$ while that of M71 $t/t_{rh} \sim 44$ (where we
have used the values of $t_{rh}$ from the Harris catalogue (2010).
Considering the difference in dynamical age, the fact that M71
populations are more mixed than those of 47 Tuc is qualitatively
consistent with the results of the simulations of the long-term
dynamical evolution of multiple-population clusters of Vesperini
et al. (2013). The simulations of Vesperini et al. (2013) also
suggest a connection between the level of spatial mixing and the
amount of mass lost due to two-body relaxation: clusters which
have lost a larger fraction of their mass are expected to be more
spatially mixed.

The extreme flattening of the stellar mass function reported by De
Marchi et al. (2007) for M71 suggests that this cluster should
indeed have suffered a significant mass loss due to two-body
relaxation. Carretta et al. (2010) also suggested, on the basis of
an excess of X-ray sources (Elsner et al. 2008), that M71 might
have lost a significant fraction of its mass. Finally, we also
point out that according to Dinescu et al. (1999) M71's orbit is
confined within 6.7 kpc from the Galactic center and the stronger
tidal field of the inner Galactic regions can enhance the
cluster's mass loss rate (see e.g. Vesperini \& Heggie 1997,
Baumgardt \& Makino 2003).

\subsection{The Aluminum Abundance in M71}

New insight on the nature of the polluters responsible for the
star-to-star variations found within Galactic GCs can be obtained
from Al, whose production through the proton-capture Mg-Al cycle
requires higher temperatures ($\geq$70MK) than that needed to
produce Na at the expenses of Ne, due to their higher Coulomb
barriers (e.g. Langer et al. 1997, Prantzos et al. 2007).
Unfortunately, studies of Al in M71 have been limited to samples
of 9-11 stars (Ram\'irez and Cohen 2002, MC09, CBG09b). These
studies have found average [Al/Fe] abundance ratios ranging from
+0.25 to +0.49, and a small scatter in Al ($\sim$0.05-0.20 dex).
As noticed by O'Connell et al. (2011) and references therein, more
metal-poor GCs tend to show larger scatter in Al. Interestingly,
the same trend described by Galactic GCs is also found in the
massive GC Omega Cen, with past studies finding a decrease in the
range of observed [Al/Fe] ratios with increasing metallicity. For
instance, an [Al/Fe] ratio as high as $\sim$+1.4 dex is found in
the more metal-poor stars of the cluster, whereas the more
metal-rich population, with a metallicity comparable to M71, only
reaches [Al/Fe]$\sim$+0.6 dex. These observational trends are
consistent with the predicted Al-yields of intermediate mass
($\sim$4-8 $M_{\odot}$) AGB stars (e.g. Ventura \& D'Antona 2009,
Karakas 2010), whose models indicate that a lower Al enhancement
is produced by more metal-rich AGBs due to the lower temperatures
in their convective envelopes and fewer dredge-up episodes.  In
particular, the [Al/Fe] ratios predicted for AGB and Super AGB
stars with metallicity comparable to M71 and M$\geq$4$M_{\odot}$,
range from 0.3-0.5 dex (Ventura et al. 2014).

We find a median [Al/Fe]=+0.37 dex (standard deviation 0.11) for
25 stars, consistent with the predictions from Ventura et al.
(2014). Using the same Al doublet, line list, and procedure, we
found that 47 Tuc has a median [Al/Fe]=+0.38 dex (standard
deviation 0.12) (Cordero et al. 2014). Despite the present-day
mass of M71 being significantly smaller than 47 Tuc's, which is
inferred from their absolute magnitudes used as proxies of their
total mass (M$_v$=-5.61 and -9.42, respectively, from Harris
1996), the clusters exhibit a similar median [Al/Fe] ratio. This
suggests that the metallicity of the polluters is responsible for
the mild Al production and modest scatter (interquartile range)
detected in these two more metal-rich clusters.

To assess what factors are important in Al production in Galactic GCs,
CBG09b have shown that a linear combination of a cluster luminosity
and metallicity correlates with the Al production (their Fig. 13).
Interestingly, M71 does not follow the trend outlined by the other
clusters of their survey. In addition to Al, M71 and 47 Tuc also show similar behavior in the light elements C
and N. As shown by Briley et al. (2001), the range and
distribution of CN and CH are identical in these two clusters,
despite their different mass, suggesting that these similarities
are a consequence of the comparable metallicity of the polluters.
Given the similar [Al/Fe] interquartile range and median value we
have found between 47 Tuc and M71, we suggest that the Al pattern
in this metallicity regime is more sensitive to the metallicity of
the polluters than to cluster mass, thus breaking the
metallicity-mass degeneracy for the Al pattern.

\begin{figure}
\includegraphics[width=0.5\textwidth]{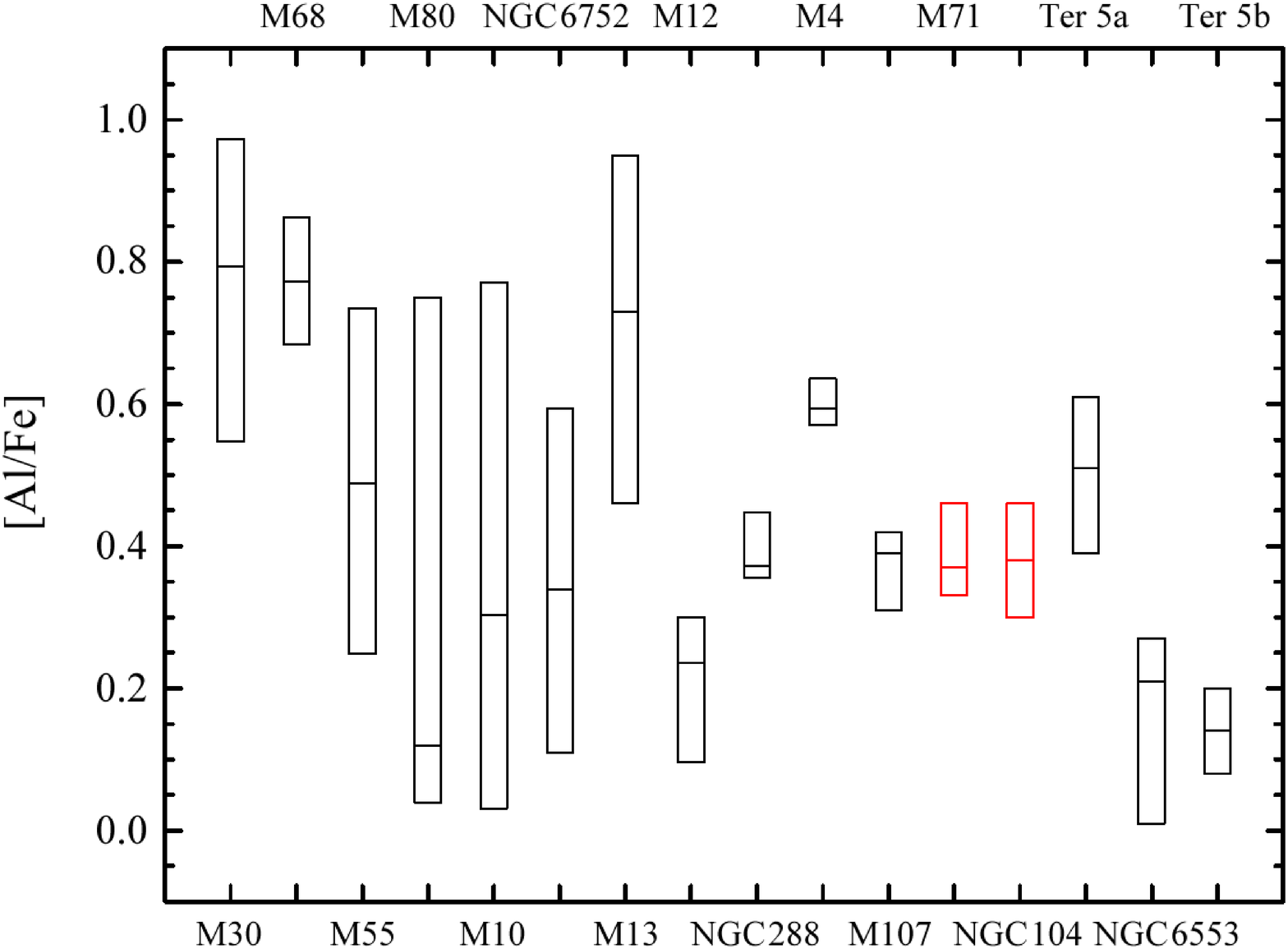}
\caption{The [Al/Fe] abundance
in M71 compared to other globular clusters arranged (from left to
right) in order of increasing metallicity. For each cluster, the
horizontal bar gives the median [Al/Fe] ratio and the lower and
upper box boundaries indicate the first and third quartiles of the
data (25\% and 75\%). Data sources are as follows: M80 from
Cavallo et al. (2004); M13 from Johnson et al. (2005); M30, M68,
M55, M10, NGC 6752, M12, NGC 288, and M4, from Carretta et al.
(2009b); M107 from O'Connell et al. (2011); 47 Tuc from Cordero et
al. (2014); Terzan 5 from Origlia et al. (2011), where a and b
represents the populations with [Fe/H]=--0.25 and [Fe/H]=+0.27,
respectively; NGC 6553 from Johnson et al. (2014; submitted). Our
results for M71 and 47 Tuc are shown in red.}
\label{fig_boxplotAl}
\end{figure}

\begin{figure}
\includegraphics[width=0.5\textwidth]{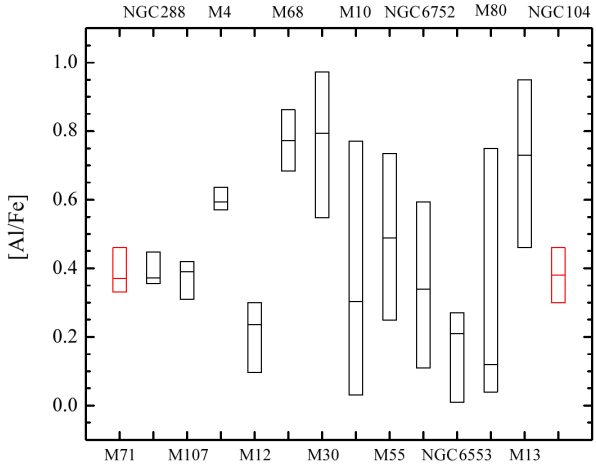}
\caption{[Al/Fe] ratios of GCs
arranged (from left to right) in order of decreasing absolute
magnitude, which trace the cluster mass. Our results for M71 and
47 Tuc are shown in red.} \label{fig_boxplotAl2}
\end{figure}

The [Al/Fe] scatter seen in M71 is consistent with other similar
metallicity clusters, as shown in Fig. \ref{fig_boxplotAl}.
Furthermore, we compare the interquartile range for the clusters
with -1.8$<$[Fe/H]$<$+0.27 (from M80 to NGC 6553) presented in
Fig. \ref{fig_boxplotAl} with the results shown in Ventura et al.
(2014). According to their Fig. 2, the maximum scatter in [Al/Fe]
produced by AGB pollution decreases monotonically as a function of
metallicity from $\sim$0.6 to $\sim$0.15 dex consistent with the
results shown in our Fig. \ref{fig_boxplotAl}. Moreover, our
interquartile range for M71 (0.13 dex) is consistent with their
prediction of a maximum scatter of 0.25 dex in [Al/Fe] at M71's
metallicity. The star-to-star scatter in [Al/Fe] decreases around
the metallicity of NCG 288 ([Fe/H]=-1.24 from CBG09a), as shown in
Fig. \ref{fig_boxplotAl}. Interestingly, a histogram of GC [Fe/H]
values suggests a bimodal distribution where the transition from
the metal-poor to metal-rich clusters occurs at roughly the same
point as the strong drop in the [Al/Fe] scatter, as seen in Fig. 6
from Muratov \& Gnedin (2010). Comparing the [Al/Fe] scatter with
GC mass, as shown in Fig. \ref{fig_boxplotAl2}, we find that the
more massive clusters exhibit a larger interquartile range for
[Al/Fe] ratios. 47 Tuc seems to be the outlier exhibiting a
smaller interquartile range within the more massive clusters,
which is probably a consequence of being a more metal-rich
cluster.

\begin{figure}
\includegraphics[width=0.5\textwidth]{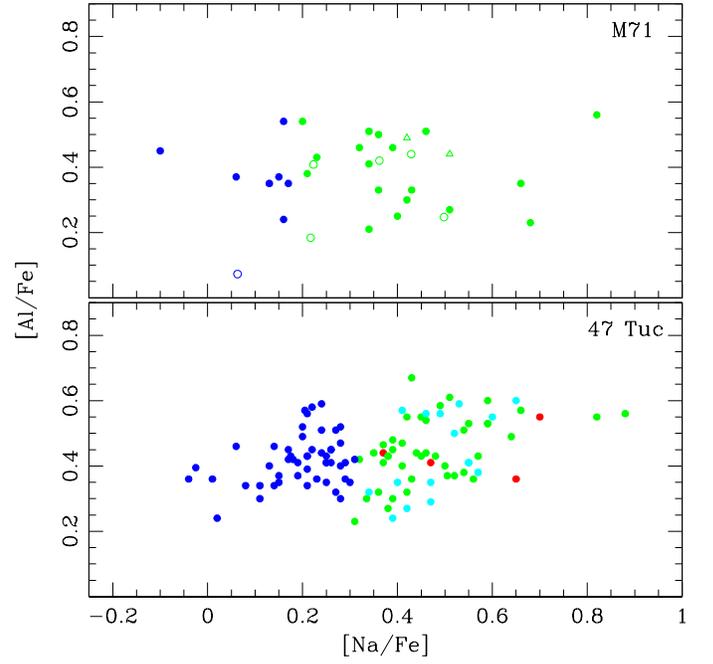}
\caption{[Al/Fe] vs. [Na/Fe].
Stars assigned to the primordial population are shown in blue,
stars assigned to the intermediate population are shown in green,
and stars belonging to the extreme population are shown in red. Stars not classified according to population are shown in cyan. The upper panel (a) shows that in our sample both the primordial
and intermediate populations exhibit a similar mean aluminum
abundance and dispersion. The lower panel (b) shows results
obtained for 47 Tuc (Cordero et al. 2014).} \label{fig_AlNa}
\end{figure}

\begin{figure}
\includegraphics[width=0.5\textwidth]{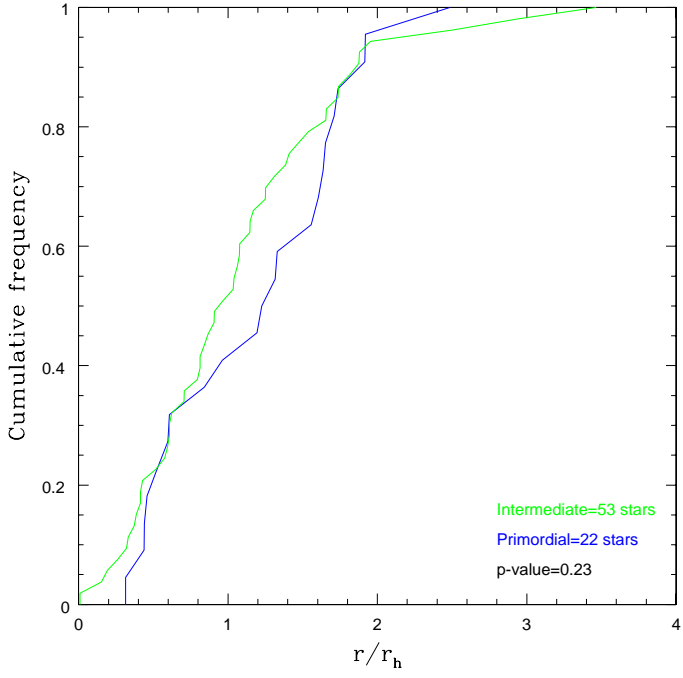}
\caption{Cumulative fraction of
stars as a function of half-mass radius for the primordial and
intermediate populations in M71. A KS test indicates that the two
populations found in M71 are not statistically different in their
radial distributions (p-value=0.23).} \label{fig_cum}
\end{figure}

The Al-Na correlation detected in metal-poor GCs (e.g. Shetrone
1996, Ivans et al. 2001, CBG09b) is not present in the more
metal-rich population of Omega Cen (e.g. Johnson \& Pilachowski
2010) or in a large sample of RGB stars of 47 Tuc (Cordero et al.
2014), although see Carretta et al. (2013) for a different
conclusion. In order to assess whether this trend is shared by the
more metal-rich GCs, we have increased the number of published
aluminum measurements in M71. No correlation of Al with Na is
found for M71 giants, as shown in Fig. \ref{fig_AlNa}.  In the
same figure we added the Al abundances for five stars from CBG09b
after removing a systematic offset of -0.15 found in a common
subsample of six stars. Additionally, we added two stars from the
sample of MC09 after correcting for a systematic offset of +0.18
based on a common subsample of seven stars. The additional Al
measurements available from the literature are consistent with the
distribution described by our sample.

The lack of correlation of Al with Na is verified by pairs of
stars within the sample that have nearly identical atmospheric
parameters but differ significantly in their Na content, while
displaying identical line strength for Al, as shown in Fig.
\ref{fig_strengths}. This visual inspection of the spectra
suggests that differences in the Al content can be subtle for
stars that belong to different generations, distinguished by their
Na abundance, and therefore Al is a less sensitive parameter to be
used to distinguish multiple generations in this metallicity
regime.

Comparing Al abundances with yields from IMAGB, FRMS, and
supermassive stars might allow to identify polluting sources.
Unfortunately, extensive predictions of [Al/Fe] yields from FRMS
or supermassive stars are not yet available.

\section{Summary}

The characterization of the Al content of M71 has stimulated the
present work, since this element can provide observational
constraints to improve our understanding of the nature of the
polluters in metal-rich GCs.

We summarize our results as the following.

1. The Fe content in our sample of 33 stars is homogeneous
(standard deviation of 0.03) while star-to-star variations
exceeding typical uncertainties were detected for O, Na, and Al.

2. We find that M71's stellar populations are more spatially mixed
than the populations of 47 Tuc. Although this result is based on a
small sample of stars and additional studies of this issue are
desirable, the difference between the level of spatial mixing in
47 Tuc and M71 is consistent with the predictions of numerical
studies of multiple-population spatial mixing (Vesperini et al.
2013).

3. Compared to more metal-poor GCs, M71 and 47 Tuc have a smaller
median [Al/Fe], consistent with model predictions for AGB stars in
this metallicity regime. The fact that these two clusters have
different total mass but the same metallicity suggests that the
metallicity of the polluters is responsible for a smaller median
[Al/Fe] and interquartile range compared to metal-poor GCs.

4. The scatter in [Al/Fe] detected in M71 follows the trend
described by more metal-rich clusters. We have studied the
variation in the [Al/Fe] scatter as a function of the cluster
metallicity and found that an abrupt change in the [Al/Fe] scatter
occurs at around the same metallicity marking the transition
between metal-poor and metal-rich GCs (e.g. Muratov \& Gnedin
2010).

5. Similarly to what found in other metal-rich clusters, no Na-Al
correlation is found in M71 (Johnson \& Pilachowski 2010; Cordero
et al. 2014).

\acknowledgments We would like to thank Karen Butler, Jena
Christensen, Charles Corson, Dianne Harmer, Jennifer Power, Heidi
Schweiker, David Summers, and Daryl Willmarth for their assistance
in the data acquisition. This research has made use of the NASA
Astrophysics Data System Bibliographic Services. This publication
makes use of data products from the Two Micron All Sky Survey,
which is a joint project of the University of Massachusetts and
the Infrared Processing and Analysis Center/California Institute
of Technology, funded by the National Aeronautics and Space
Administration and the National Science Foundation. This work was supported by Sonderforschungsbereich SFB 881 ``The Milky Way System'' (subprojects A4 and A5) of the German Research Foundation (DFG). CAP
acknowledges the generosity of the Kirkwood Research Fund at
Indiana University. CIJ gratefully acknowledges support from the Clay Fellowship,
administered by the Smithsonian Astrophysical Observatory. Enrico Vesperini acknowledges supports from
grant NASA-NNX13AF45G. We thank the anonymous referee for her/his suggestions in improving this paper.

\clearpage

\clearpage

\begin{table}
\begin{center}
\begin{tabular}{ccc}
\hline\hline
UT Date  & Exp. Time & Wavelength  \\
         & (s) & ($\AA$)\\
\hline
25 May 2000  & 3600 & 6490-6800 \\
19 June 2000  & 3600 & 6490-6800 \\
10 July 2000  & 3600 & 6490-6800 \\
19 July 2003  & 3600 & 6490-6800 \\
22 June 2013 & 2700 & 6100-6400 \\
23 June 2013  & 2700 & 6100-6400 \\
3  July 2013  & 2700 & 6100-6400 \\
4  July 2013  & 2700 & 6100-6400 \\
21 July 2013  & 2700 & 6100-6400 \\
\hline\hline
\end{tabular} 
\caption{Log of M71 Observations\label{table_log}}
\end{center} 
\end{table}

\clearpage

\begin{sidewaystable}
{\tiny
\begin{tabular}{lccccccccccccccc}
\hline\hline
 & R.A. & DEC & V & K & T$_{\mbox{eff}}$ & & & $v_t$ & & & & & $RV_H$ & & r \\
Star & J2000 & J2000 & & & (K) & log g & [Fe/H] & (km s$^{-1}$) & [O/Fe] & [Na/Fe]
 & [Al/Fe] & Prob & (km s$^{-1}$) & Pop & (arcmin) \\
 \hline
1-88  &   298.4367500 &   18.7636000  &   14.26   &   11.800    &   5175    &   $  2.44    $    &   $ -0.82   $ &   $ 1.65    $ &     $\ldots$   &   $\phantom{-}   0.52    $    &     $\ldots$   &   $  90  $    &   $ -25.0   $ &     I     &   $  1.02    $\\
1-101       &   298.4373750 &   18.7721583  &   14.39   &   11.850   &   5295    &   $  2.45    $    &   $ -0.81   $ &   $ 1.50    $ &   0.37    &   $\phantom{-}   0.35    $    &     $\ldots$   &   $  90  $    &   $ -19.5   $ &     I     &   $  0.55    $\\
KC-118      &   298.4189167 &   18.8189389  &   14.55   &   11.906  &   5170    &   $  2.47    $    &   $ -0.83   $ &   $ 1.50    $ &     $\ldots$   &   $\phantom{-}   0.25    $    &     $\ldots$   &   $  95  $    &   $ -23.8   $ &     I     &   $  2.77    $\\
X       &   298.4212917 &   18.7365472  &   14.48   &   11.814  &   5145    &   $  2.43    $    &   $ -0.83   $ &   $ 1.50    $ &     $\ldots$   &   $\phantom{-}   0.17    $    &   0.35    &   $  90  $    &   $ -20.8   $ &     P     &   $  2.86    $\\
1-50  &   298.4739167 &   18.8117278  &   14.53   &   11.860   &   5140    &   $  2.44    $    &   $ -0.82   $ &   $ 1.60    $ &     $\ldots$   &   $\phantom{-}   0.16    $    &   0.24    &   $  92  $    &   $ -26.8   $ &     P     &   $  2.60    $\\
1-33  &   298.4602083 &   18.7849833  &   14.40    &   11.724  &   5130    &   $  2.39    $    &   $ -0.84   $ &   $ 1.50    $ &   0.62    &   $\phantom{-}   0.18    $    &     $\ldots$   &   $  91  $    &   $ -19.4   $ &     P     &   $  1.00    $\\
1-107       &   298.4286667 &   18.7771611  &   13.70    &   10.966  &   5062    &   $  2.08    $    &   $ -0.79   $ &   $ 1.60    $ &     $\ldots$   &   $\phantom{-}   0.33    $    &     $\ldots$   &   $  90  $    &   $ -23.8   $ &     I     &   $  0.86    $\\
1-62  &   298.4191250 &   18.8079833  &   14.56   &   11.757  &   4985    &   $  2.39    $    &   $ -0.83   $ &   $ 1.50    $ &     $\ldots$   &   $  -0.10   $    &   0.45    &   $  83  $    &   $ -23.2   $ &     P     &   $  2.22    $\\
1-75  &   298.4231667 &   18.7953722  &   14.85   &   11.940   &   4870    &   $  2.45    $    &   $ -0.86   $ &   $ 1.50    $ &     $\ldots$   &   $\phantom{-}   0.62    $    &     $\ldots$   &   $  94  $    &   $ -21.5   $ &     I     &   $  1.52    $\\
1-3   &   298.4448333 &   18.7502389  &   14.93   &   12.014  &   4870    &   $  2.47    $    &   $ -0.84   $ &   $ 1.50    $ &   0.62    &   $\phantom{-}   0.34    $    &   0.51    &   $  95  $    &   $ -24.2   $ &     I     &   $  1.74    $\\
1-111       &   298.4017083 &   18.7690222  &   14.89   &   11.966  &   4850    &   $  2.05    $    &   $ -0.81   $ &   $ 1.40    $ &     $\ldots$   &   $\phantom{-}   0.36    $    &   0.33    &   $  88  $    &   $ -21.6   $ &     I     &   $  2.46    $\\
1-58  &   298.4383333 &   18.8107361  &   14.93   &   12.000  &   4845    &   $  2.47    $    &   $ -0.78   $ &   $ 1.50    $ &     $\ldots$   &   $\phantom{-}   0.55    $    &     $\ldots$   &   $  94  $    &   $ -24.3   $ &     I     &   $  1.92    $\\
KC-130      &   298.3927500 &   18.7820667  &   14.30    &   12.090   &   4700    &   $  2.15    $    &   $ -0.82   $ &   $ 1.40    $ &   0.26    &   $\phantom{-}   0.16    $    &   0.54    &   $  94  $    &   $ -19.0   $ &     P     &   $  2.90    $\\
1-14  &   298.4796250 &   18.7596417  &   13.86   &   10.763  &   4680    &   $  1.95    $    &   $ -0.87   $ &   $ 1.50    $ &   0.53    &   $\phantom{-}   0.39    $    &   0.46    &   $  93  $    &   $ -24.0   $ &     I     &   $  2.35    $\\
1-51  &   298.4633333 &   18.8086583  &   14.34   &   11.220   &   4660    &   $  2.13    $    &   $ -0.76   $ &   $ 1.50    $ &     $\ldots$   &   $\phantom{-}   0.20    $    &   0.54    &   $  87  $    &   $ -22.9   $ &     I     &   $  2.09    $\\
1-68  &   298.4335833 &   18.7863028  &   14.38   &   11.240   &   4640    &   $  2.14    $    &   $ -0.84   $ &   $ 1.45    $ &   0.47    &   $\phantom{-}   0.21    $    &   0.38    &   $  91  $    &   $ -19.9   $ &     I     &   $  0.72    $\\
1-1   &   298.4683333 &   18.7484556  &   14.13   &   10.950   &   4605    &   $  2.01    $    &   $ -0.83   $ &   $ 1.40    $ &   0.16    &   $\phantom{-}   0.45    $    &     $\ldots$   &   $  94  $    &   $ -23.8   $ &     I     &   $  2.31    $\\
1-9   &   298.4544583 &   18.7668250  &   13.43   &   10.186  &   4550    &   $  1.70    $    &   $ -0.81   $ &   $ 1.70    $ &     $\ldots$   &   $\phantom{-}   0.43    $    &   0.33    &   $  95  $    &   $ -20.4   $ &     I     &   $  0.96    $\\
1-95  &   298.4209583 &   18.7683194  &   13.39   &   10.120   &   4530    &   $  1.67    $    &   $ -0.83   $ &   $ 1.70    $ &   0.31    &   $\phantom{-}   0.40    $    &   0.25    &   $  94  $    &   $ -21.1   $ &     I     &   $  1.45    $\\
1-56  &   298.4511667 &   18.8071194  &   13.25   &   9.873   &   4445    &   $  1.56    $    &   $ -0.80   $ &   $ 1.75    $ &   0.35    &   $\phantom{-}   0.42    $    &   0.3 &   $  96  $    &   $ -20.3   $ &     I     &   $  1.73    $\\
KC-22       &   298.4133333 &   18.7342639  &   13.97   &   10.544  &   4405    &   $  1.83    $    &   $ -0.83   $ &   $ 1.55    $ &   0.56    &   $\phantom{-}   0.15    $    &   0.37    &   $  87  $    &   $ -24.1   $ &     P     &   $  3.20    $\\
1-21 &   298.4480000 &   18.7714667  &   13.02   &   9.588   &   4400    &   $  1.44    $    &   $ -0.79   $ &   $ 1.80    $ &   0.43    &   $\phantom{-}   0.06    $    &   0.37    &   $  95  $    &   $ -21.2   $ &     P     &   $  0.52    $\\
S232        &   298.4902917 &   18.7992389  &   13.56   &   10.042  &   4340    &   $  1.62    $    &   $ -0.84   $ &   $ 1.55    $ &   0.23    &   $\phantom{-}   0.66    $    &   0.35    &   $  93  $    &   $ -26.4   $ &     I     &   $  2.91    $\\
1-53  &   298.4609583 &   18.8188167  &   12.97   &   9.271   &   4220    &   $  1.30    $    &   $ -0.84   $ &   $ 1.65    $ &   0.22    &   $\phantom{-}   0.82    $    &   0.56    &   $  95  $    &   $ -24.7   $ &     I     &   $  2.57    $\\
I       &   298.4365833 &   18.7764722  &   12.38   &   8.663   &   4210    &   $  1.05    $    &   $ -0.84   $ &   $ 1.60    $ &   0.27    &   $\phantom{-}   0.46    $    &   0.51    &   $  93  $    &   $ -29.6   $ &     I     &   $  0.44    $\\
1-64  &   298.4423333 &   18.7906639  &   13.13   &   9.395   &   4195    &   $  1.34    $    &   $ -0.83   $ &   $ 1.60    $ &   0.44    &   $\phantom{-}   0.36    $    &   0.5 &   $  95  $    &   $ -17.5   $ &     I     &   $  0.69    $\\
S       &   298.4161667 &   18.7314778  &   13.01   &   9.217   &   4160    &   $  1.26    $    &   $ -0.77   $ &   $ 1.70    $ &   0.05    &   $\phantom{-}   0.34    $    &   0.21    &   $  81  $    &   $ -21.8   $ &     I     &   $  3.26    $\\
A9      &   298.4061250 &   18.7500139  &   13.10    &   9.052   &   4025    &   $  1.19    $    &   $ -0.83   $ &   $ 1.55    $ &   0.36    &   $\phantom{-}   0.13    $    &   0.35    &   $  86  $    &   $ -24.5   $ &     P     &   $  2.76    $\\
A4      &   298.4720000 &   18.7798278  &   12.28   &   8.040    &   3935    &   $  0.77    $    &   $ -0.84   $ &   $ 1.65    $ &   0.33    &   $\phantom{-}   0.18    $    &   0.41    &   $  94  $    &   $ -21.3   $ &     P     &   $  1.61    $\\
1-45  &   298.4511250 &   18.8007056  &   12.35   &   8.094   &   3930    &   $  0.79    $    &   $ -0.76   $ &   $ 1.50    $ &   0.11    &   $\phantom{-}   0.68    $    &   0.23    &   $  94  $    &   $ -19.4   $ &     I     &   $  1.36    $\\
1-46  &   298.4646250 &   18.8017222  &   12.34   &   7.968   &   3880    &   $  0.74    $    &   $ -0.82   $ &   $ 1.50    $ &   0.52    &   $\phantom{-}   0.32    $    &   0.46    &   $  93  $    &   $ -23.2   $ &     I     &   $  1.80    $\\
1-113       &   298.4406250 &   18.7986778  &   12.48   &   7.974   &   3825    &   $  0.73    $    &   $ -0.81   $ &   $ 1.50    $ &   0.31    &   $\phantom{-}   0.23    $    &   0.43    &   $  95  $    &   $ -21.3   $ &     I     &   $  1.18    $\\
H       &   298.4477917 &   18.7809472  &   12.08   &   7.449   &   3780    &   $  0.51    $    &   $ -0.77   $ &   $ 1.50    $ &   0.05    &   $\phantom{-}   0.51    $    &   0.27    &   $  82  $    &   $ -21.3   $ &     I     &   $  0.25    $\\
\hline\hline
\end{tabular} 
\caption{Atmospheric Parameters and Abundances.  \label{tabshort}}
}
\end{sidewaystable}

\clearpage

\begin{table}
\begin{center}
\begin{tabular}{rrrrrrrr}
\hline\hline
Ion & $T_{\mbox{eff}}\pm100$ & $\log g\pm0.20$ & [M/H]$\pm$0.10 & $v_t\pm$0.20 & No. & $\sigma/\sqrt N$ & $\sigma_{\mbox{total}}$  \\
 & (K) & (cgs)& (dex)& (km s$^{-1}$) & lines & (dex) & (dex) \\
 \hline
1-64\\
Fe I &$  0.03    $&$ 0.03    $&$ -0.10    $&$ 0.02    $&$ 36    $&$ 0.02    $&$ 0.11    $\\
O   &$  0.01    $&$ 0.06    $&$ -0.04   $&$ 0.01    $&$  1   $&$ 0.07    $&$ 0.10    $\\
Na  &$  0.03    $&$ -0.02   $&$ -0.07   $&$ -0.05   $&$ 2    $&$ 0.07    $&$ 0.12    $\\
Al  &$  0.05    $&$ -0.01   $&$ -0.05   $&$ -0.03   $&$ 2    $&$
0.07    $&$ 0.10    $\\\hline
X\\
Fe I &$  0.10 $&$ 0.00   $&$ -0.07   $&$ -0.01   $&$  23   $&$ 0.03    $&$ 0.13    $\\
Na  &$  -0.03   $&$ -0.06   $&$ -0.04   $&$ -0.02   $&$  2   $&$ 0.07    $&$ 0.11    $\\
Al  &$  0.03    $&$ -0.04   $&$ -0.03   $&$ -0.04   $&$  2   $&$ 0.07    $&$ 0.10    $\\
\hline\hline
\end{tabular} 
\caption{Abundance Sensitivity to Model Atmosphere
Parameters. \label{table_sensitivities}}
\end{center} 
\end{table}

\end{document}